\documentclass{JHEP3}
\usepackage{amssymb,amsmath}
\usepackage{epsfig}
\usepackage{citesort}

\title{Exploiting scale dependence in cosmological averaging}

\author{Teppo Mattsson$^{1,2,}$\footnote{E-mail: teppo.mattsson@helsinki.fi} ~and Maria Ronkainen$^{1,2,}$\footnote{E-mail: maria.ronkainen@helsinki.fi}\\
${}^1$ Helsinki Institute of Physics, P.O. Box 64, FIN-00014 University of Helsinki, Finland\\
${}^2$ Department of Physical Sciences, P.O. Box 64, FIN-00014
University of Helsinki, Finland}

\abstract{We study the role of scale dependence in the Buchert
averaging method, using the flat Lemaitre-Tolman-Bondi model as a
testing ground. Within this model, a single averaging scale gives
too coarse predictions, but by replacing it with the distance of the
objects $R(z)$ for each redshift $z$, we find an
${\mathcal{O}}(1\%)$ precision at $z < 2$ in the averaged
luminosity and angular diameter distances compared to their exact expressions.
At low redshifts, we show the improvement for generic inhomogeneity profiles,
and our numerical computations further verify it up to redshifts $z\sim 2$.
At higher redshifts, the method breaks down due to its inability to
capture the time evolution of the inhomogeneities. We also
demonstrate that the running smoothing scale $R(z)$ can mimic acceleration,
suggesting it could be at least as important as the backreaction in
explaining dark energy as an inhomogeneity induced illusion.}

\preprint{HIP-2007-45/TH}


\begin{document}

\section{Introduction}\label{intro}

The current cosmological observations seem to get the simplest
and rather concordant interpretation in the homogeneous and
isotropic expanding universe models, with late-time acceleration
starting around the redshift $z \sim 0.6$
\cite{Riess:2004nr,Eisenstein:2005su,Spergel:2006hy}. The
acceleration is usually seen as an evidence for dark energy, most
often in the form of a cosmological constant or vacuum energy.
However, the enormous fine-tuning needed to explain both the size
and the timing of such an energy component has raised serious doubts
about its correctness and thus justified the search for alternatives
\cite{Copeland:2006wr,Straumann:2006tv,Sahni:2006pa,Blanchard:2003du,Hunt:2004vt,Hunt:2007dn}.

Perhaps the most natural alternative explanation so far has arisen
from the inhomogeneous cosmological models. The point in these
models is that suitable inhomogeneities can have a similar effect on
the observations of light as accelerating expansion in the
homogeneous models
\cite{PascualSanchez:1999zr,Celerier:1999hp,Tomita:2000jj,Iguchi:2001sq}.
Although gained more popularity only recently, the actual idea is
not a new one (see \cite{Krasinski,Plebanski:2006sd}). Indeed, already the
pioneers of cosmology were careful to point out the potential
inadequacy of the simplest homogeneous models in describing the real
universe \cite{Lemaitre:1933qe,Tolman:1934za,Bondi:1947av}.

Two conceptually rather different kind of inhomogeneities have been
proposed as the culprit for the apparent acceleration. Firstly, the
non-perturbative effects of the well-established lumpiness of
galaxies and galaxy clusters are still unknown, and could
potentially mimic acceleration \cite{Rasanen:2006kp}. A virtue in
this scenario is that it would connect the growth of nonlinear
structure with the start of the acceleration era and thus solve the
coincidence problem \cite{Schwarz:2002ba}. Secondly, there
are inhomogeneities also on scales larger than the observed clustering.
Indeed, the increasing accuracy of the cosmological observations has revealed larger and larger
voids \cite{Rudnick:2007kw,Tikhonov:2007di}. For these kind of smooth
large scale inhomogeneities, there are exactly solvable models, such
as the LTB class of solutions, which can mimic acceleration without
dark energy \cite{Celerier:1999hp,Iguchi:2001sq}. However, due to
the complexity of the Einstein field equations, there are no exact
solutions for the small scale lumpiness of the universe. Thus,
some level of coarse graining has to be introduced, which has
usually been done in the form of an averaging method \cite{Ellis:2005uz}. One
of the most popular methods in cosmology is the Buchert averaging
\cite{Buchert:1999er,Buchert:2007ik}, which is also the one
considered in this work.

The conventional way to apply the Buchert formalism is to average over a
single domain, larger than the supposed scale of statistical homogeneity in the galaxy distribution.
Consequently, it has been speculated that the averaging method fails if there are inhomogeneities at
large scales as well \cite{Rasanen:2006kp}. Motivated by the recently observed large voids and
superclusters, we try to extend the averaging method to work with large scale inhomogeneities
by going beyond the single scale approximation. In fact, the point we want to bring out is
that it would be physically reasonable to
replace the single scale $R$ by the distance of the objects $R(z)$ for
each redshift $z$, since the distance the observed light travels depends on how far the object is.

A commonly presented conjecture in cosmological averaging is that
the backreaction of inhomogeneities causes the acceleration of the
average expansion and could thus account for the observations
\cite{Rasanen:2006kp,Buchert:2007ik,Paranjape:2006cd}. However, as demonstrated in this work,
by promoting the single averaging scale $R$ to the redshift dependent
function $R(z)$, it is possible to mimic acceleration even in the
absence of backreaction. Indeed, at least within the flat LTB model, this extension gives rise to a definite
improvement in accuracy of the averaged luminosity and angular
diameter distances compared to the single scale case, suggesting
that the averaging method can also be utilized for large scale inhomogeneities.
Naturally, in the case a single scale would suffice, the running scale approach
reduces to give the same predictions as the conventional single scale approach. An
additional virtue is the computational simplicity of this generalization.

The paper is organized as follows. In Sect.\ \ref{buchertLTB}, we
derive the observable distance-redshift relations for the averaged
LTB model with flat spatial sections. The principal results are
presented in Sect.\ \ref{scale}, where we compare these relations with
their exact counterparts, using various implementations of the
running averaging scale $R(z)$. For general inhomogeneity profiles,
we utilize power series to make an analytic comparison at low
redshifts. At higher redshifts, we employ numerical computations for
two explicit profiles: a bubble inhomogeneity that fits the
supernova observations and periodic inhomogeneities as a toy model
for structure. The results of the different cases are discussed in
Sect.\ \ref{discussion}. Finally, Sect.\ \ref{conclusions} contains
our conclusions.

\section{The Buchert equations for the flat LTB model}\label{buchertLTB}

In this section, we calculate the Buchert equations for the
spatially flat, spherically symmetric LTB model with pressureless
matter as the only source. The backreaction of the model vanishes
identically \cite{Paranjape:2006cd} and hence the only difference from the
homogeneous and flat matter dominated FRW case is the scale
dependence of the averaged quantities. This property makes the model
especially useful in extracting the effect of the averaging scale on
the observable quantities, such as the relation of redshift to the
angular diameter distance and to the luminosity distance.

\subsection{Observations in the LTB model}\label{flatLTB}

The line-element of the spatially flat LTB model with the spatial
origin at the symmetry center reads as
\begin{equation}\label{LTBmetric}
ds^2 = - dt^2 + (A'(r,t))^2 dr^2 + A^{2}(r,t) \left(
d\theta^2 + \sin^2\theta d\varphi^2 \right)~,
\end{equation}
where $A(r,t)$ is the scale function having both temporal and
spatial dependence, and we use the following shorthand notations for
the partial derivatives: $'\equiv \frac{\partial}{\partial r}$ and
$\dot{} \equiv \frac{\partial}{\partial t}$. This metric was first
studied by Lemaitre \cite{Lemaitre:1933qe}, Tolman
\cite{Tolman:1934za} and Bondi \cite{Bondi:1947av}; later, it has
been used in various astronomical and cosmological contexts
\cite{Krasinski,Plebanski:2006sd}. Although commonly called a toy model,
the metric (\ref{LTBmetric}) is an exact solution of the Einstein
equations and the perfectly homogeneous FRW model is only a special
case of it, obtained in the limit: $A(r,t) \rightarrow a(t)r$, where
$a(t)$ is the FRW scale factor. Our notation and parametrization
follows Ref. \cite{Enqvist:2006cg}.

The Einstein equations for the metric (\ref{LTBmetric}) reduce to
the generalized Friedmann equation
\begin{equation}\label{FlatFriidman}
H(r,t) = H_0(r) \left(\frac{A_0(r)}{A(r,t)} \right)^{3/2} ~,
\end{equation}
where $H(r,t) \equiv \dot{A}(r,t)/A(r,t)$ is the LTB version of the
Hubble function, $A_0(r) \equiv A(r,t_0)$ is the scale function at a
reference time $t_0$, $H_0(r) \equiv H(r,t_0)$ is the position
dependent Hubble constant, and to the time evolution equation of the
matter density
\begin{equation}\label{rhotime}
\rho_M (r,t) = \frac{3H_0^2(r)}{8 \pi G}  \left[1 +
\frac{2 A_0(r) H_0'(r)}{3A_0'(r) H_0(r)}\right] \left(
\frac{A_0^2(r)A_0'(r)}{A^2(r,t)A'(r,t)} \right)~.
\end{equation}
The integration of Eq.\ (\ref{FlatFriidman}) w.r.t.\ time yields
\begin{equation}\label{Asolved}
\frac{A(r,t)}{A_0(r)} = \left[1+ \frac{3}{2} H_0(r) (t-t_0) \right]^{2/3}~.
\end{equation}
Substituting Eq.\ (\ref{Asolved}) in Eq.\ (\ref{FlatFriidman}) gives
the time evolution of the Hubble function:
\begin{equation}\label{Timedependenthubble}
H(r,t) = \frac{H_0(r)}{1 + \frac{3H_0(r)}{2}(t-t_0)}~.
\end{equation}
The choice of $A_0(r)$ represents a coordinate freedom, similar to
the normalization of $a(t_0)$ in the FRW case; here we set
$A_0(r)=r$. Consequently, the flat LTB model is uniquely determined by the
free function $H_0(r)$. Plugging Eq.\ (\ref{Asolved}) in Eq.\
(\ref{rhotime}) then gives the explicit time dependence of the
matter distribution:
\begin{equation}\label{rhotime2}
\rho_M(r,t) = \frac{3H_0^2(r) + 2r H_0'(r)H_0(r)}{8\pi
G[1+\frac{3H_0(r)}{2}(t-t_0)][r(t-t_0)H_0'(r)+ (1+
\frac{3H_0(r)}{2}(t-t_0))]}~.
\end{equation}

Eqs.\ (\ref{Timedependenthubble}) and (\ref{rhotime2}) show that the
inhomogeneities in the flat LTB model correspond to "decaying
modes"\ \cite{Silk}, i.e.\ the model evolves towards homogeneity at
late times. This is just the opposite to the expected structure
formation in the real universe, but for low-redshift observations,
this should not make a difference. In the nonflat LTB model with
simultaneous Big Bang, the inhomogeneities correspond to "growing
modes"\, and in this respect, that model would be more realistic.
However, as the backreaction does not vanish in the nonflat model,
it would be harder to distinguish the effect of the running
smoothing scale from the backreaction effects, which is why we use
the flat LTB metric (\ref{LTBmetric}). The price to pay is that, for
most of the inhomogeneity profiles $H_0(r)$, we have to restrict our
considerations to observations at low redshifts ($z \lesssim 2$).

To study observable properties of light, we need the pair of radial
geodesic equations determining the relations between the coordinates
and the observable redshift, $t(z)$ and $r(z)$, given by
\cite{Bondi:1947av}
\begin{equation}\label{dtdz}
\frac{dt}{dz} = \frac{-A'(r,t)}{(1+z) \dot{A}'(r,t)}~,
\end{equation}
\begin{equation}\label{drdz}
\frac{dr}{dz} =
\frac{1}{(1+z) \dot{A}'(r,t)}~.
\end{equation}
To calculate the right hand sides of Eqs.\ (\ref{dtdz}) and
(\ref{drdz}), we need the following derivatives of the scale function
(\ref{Asolved}):
\begin{equation}\label{Aprime}
A'(r,t)= \left(1+\frac{3}{2} (t-t_0) H_0(r)\right)^{\frac{2}{3}}+r (t-t_0)
H_0'(r)\left(1+\frac{3}{2} (t-t_0) H_0(r)\right)^{-\frac{1}{3}}~,
\end{equation}
\begin{equation*}
\dot{A}'(r,t)=\left(1+\frac{3}{2} (t-t_0)
H_0(r)\right)^{-\frac{4}{3}} \bigg[ H_0(r) \left(1+\frac{3}{2} (t-t_0)
H_0(r)\right)+
\end{equation*}
\begin{equation}\label{Adotprime}
+r H_0'(r) \bigg(1 + (t-t_0) H_0(r) \bigg) \bigg]~.
\end{equation}

The relation of the redshift to the energy flux $F$, or the
luminosity-distance, defined as $d_L \equiv \sqrt{{L}/{4 \pi F}}$
with $L$ the total power radiated by the source, as well as to the
angular-diameter distance $d_A$ are given by \cite{Ellis}
\begin{equation}\label{lumdist}
d_L (z) = (1+z)^2 A(r(z),t(z)) ~,
\end{equation}
\begin{equation}\label{angdist}
d_A (z) = A(r(z),t(z)) ~.
\end{equation}
As the relations $t(z)$ and $r(z)$ are determined by Eqs.\
(\ref{dtdz}), (\ref{drdz}), (\ref{Aprime}), (\ref{Adotprime}), and
the scale function $A(r,t)$ by Eq.\ (\ref{Asolved}), using Eqs.\
(\ref{lumdist}) and (\ref{angdist}), one can compute the observables
$d_L(z)$ and $d_A(z)$ for a given $z$. In Sect.\ \ref{scale}, we
compare the exact forms of these observable relations to their
Buchert-averaged counterparts with various implementations of the
scale dependence and different inhomogeneity profiles $H_0(r)$. For
this, we need to first apply the Buchert formalism to the flat LTB
metric (\ref{LTBmetric}).

\subsection{Observations in the averaged LTB model}\label{puhhert}

To construct a coarse grained description of the inhomogeneous
universe, one usually has to average dynamical quantities of
Einstein's gravitation theory (see \cite{Ellis:2005uz} for a
review). It seems physically more correct to first calculate the
Einstein field $\mathbf{G}(\mathbf{g})$ for the exact metric
$\mathbf{g}$ and only then average $\langle \mathbf{G}(\mathbf{g})
\rangle$, than to calculate the Einstein field for the averaged
metric $\mathbf{G} (\langle \mathbf{g} \rangle)$. The reason is that
the Einstein field is more closely related to physical quantities
whereas the metric corresponds to gravitational potentials, whose
derivatives determine the physics. Since in Einstein's gravity the
field $\mathbf{G}$ depends nonlinearly on the metric $\mathbf{g}$,
its evaluation does not commute with averaging: $\langle
\mathbf{G}(\mathbf{g}) \rangle \neq \mathbf{G} (\langle \mathbf{g}
\rangle )$. Hence the issue is not only a conceptual one, but in
general leads to physically different predictions. However, in the
absence of nonlinear inhomogeneities the two approaches lead to
identical results. In fact, the standard model of cosmology builds
on the assumption that the undoubtedly existent, intense small scale
lumpiness has no cosmological significance. In any case, the
inadequacy of the standard model to explain the observations without
a severely fine-tuned cosmological constant should, at the very least,
justify the more thorough considerations of this assumption.

By averaging the scalar part of the Einstein equations
in the above explained order, one arrives at
the Buchert equations describing the averaged dynamics of a general
irrotational dust universe \cite{Buchert:1999er}:
\begin{eqnarray}
3\frac{\ddot{a}_\mathcal{D}}{a_\mathcal{D}} &=& -4 \pi G \langle\rho\rangle_\mathcal{D}+\mathcal{Q}_\mathcal{D} \label{Buchert1} \\
3\left(\frac{\dot{a}_\mathcal{D}}{a_\mathcal{D}}\right)^2 &=& 8 \pi
G\langle\rho\rangle_\mathcal{D}-\frac{1}{2}\langle
{}^{(3)}R\rangle_\mathcal{D}-\frac{1}{2}\mathcal{Q}_\mathcal{D}
\label{Buchert2} \\
\frac{\partial}{\partial t} \langle\rho\rangle_\mathcal{D}&=&-3\frac{\dot{a}_\mathcal{D}}{a_\mathcal{D}}\langle\rho\rangle_\mathcal{D}~,
\label{Buchert3}
\end{eqnarray}
where $\rho$ is the matter density and the difference between the
Buchert acceleration equation (\ref{Buchert1})
and its homogeneous FRW counterpart is known as the backreaction
\begin{equation}\label{backreaction}
\mathcal{Q}_\mathcal{D} (t) \equiv\frac{2}{3}(\langle\theta^2\rangle_\mathcal{D}-\langle\theta\rangle^2_\mathcal{D})- \langle\ \hspace{-3pt} \sigma^{\mu \nu}\sigma_{\mu \nu}\rangle_\mathcal{D}~,
\end{equation}
where $\sigma^{\mu \nu} \sigma_{\mu \nu} \geq 0$ represents the
shear,
\begin{equation}\label{effscalefactor}
a_\mathcal{D}(t)\equiv\left( \frac{\int_{\mathcal{D}}\sqrt{{\rm
det}[g_{ij}]}d^3x}{\int_{\mathcal{D}}\sqrt{{\rm
det}[g_{ij}(t=t_0)]}d^3x}\right)^{\frac{1}{3}}~
\end{equation}
is the averaged scale factor, $\mathcal{D}$ is the averaging
domain, $g_{ij}$ is the spatial part of the metric, ${}^{(3)}R$ is the
curvature scalar of the $t={\rm{const.}}$ spatial hypersurfaces,
$\theta \equiv \nabla_\mu u^\mu$ is the expansion scalar and the
spatial average of a scalar $S$ is defined as
\begin{equation}\label{average}
\langle S \rangle_{\mathcal{D}} (t) \equiv \frac{\int_{\mathcal{D}} S( x^i , t) \sqrt{ {\rm{det}}
[g_{ij}]} d^3 x}{\int_{\mathcal{D}} \sqrt{ {\rm{det}} [g_{ij}]} d^3 x }~.
\end{equation}

The average expansion accelerates if the right hand side of Eq.\
(\ref{Buchert1}) is positive; this can be achieved by having large
enough variance of the expansion rate although it is partially
counterbalanced by the average shear. The variance gets large when
contracting ($\theta<0$) and expanding ($\theta>0$) regions coexist,
and in fact the average acceleration has been sometimes connected to
gravitational collapse
\cite{Rasanen:2006kp,Apostolopoulos:2006eg,Kai:2006ws}. Anyhow, as
demonstrated in Ref. \cite{Paranjape:2006cd}, a globally expanding
dust universe can have average acceleration as well.

In the context of the averaged universe, whose dynamics is given by
Eqs.\ (\ref{Buchert1}), (\ref{Buchert2}) and (\ref{Buchert3}), it is
natural to further assume that the average metric takes the FRW
form:
\begin{equation}\label{averagedmetric}
ds^2 = -dt^2 + a_{\mathcal{D}}^2(t) \left[ \frac{dr^2}{1-k_{\mathcal{D}} r^2} + r^2 (d\theta^2 + \sin^2\theta d\varphi^2) \right] ~,
\end{equation}
as was recently highlighted by Paranjape and Singh \cite{Paranjape:2006ww} (see also \cite{Buchert:2007ik}). Although the form of the
metric (\ref{averagedmetric}) is the same as in the perfectly homogeneous universe, the time evolution
of the scale factor $a_{\mathcal{D}}(t)$ and the spatial curvature $k_{\mathcal{D}}$ are in general
different from any FRW model.

Recalling the definition of the shear tensor
\begin{equation}\label{shear}
\sigma_{\mu \nu} \equiv \frac{1}{2} ( \nabla_\mu u_\nu + \nabla_\nu
u_\mu ) - \frac{1}{3} ( g_{\mu \nu} + u_\mu u_\nu) \nabla_\alpha
u^\alpha ~ ,
\end{equation}
where $u^\mu$ is the four-velocity of the dust, we obtain the shear
and the expansion scalars of the LTB metric in the coordinates of
Eq.\ (\ref{LTBmetric}):
\begin{equation}\label{shear2}
\sigma^{\mu \nu} \sigma_{\mu \nu} = \frac{2}{3} \left(
\frac{\dot{A}(r,t)}{A(r,t)} - \frac{\dot{A}'(r,t)}{A'(r,t)}
\right)^2~,
\end{equation}
\begin{equation}\label{expscalar}
\theta = 2 \frac{\dot{A}(r,t)}{A(r,t)} +
\frac{\dot{A}'(r,t)}{A'(r,t)} \equiv 2H(r,t) + H_r(r,t)~.
\end{equation}
Using the field equation (\ref{FlatFriidman}), one finds that the
two Hubble functions $H(r,t)$ and $H_r(r,t)$, defined via Eq.\
(\ref{expscalar}), are related at the reference time $t=t_0$ as
\begin{equation}\label{hubbler}
H_r(r,t_0) = H_0(r) + r H_0'(r)~,
\end{equation}
so that the expressions for the shear and the expansion rate simplify to:
\begin{eqnarray}\label{shearnow}
\sigma^{\mu \nu} \sigma_{\mu \nu}(t=t_0) &=& \frac{2}{3} (rH_0'(r))^{2}~, \\
\theta(t=t_0) &=& 3 H_0(r) + rH_0'(r)\label{expansionnow}~.
\end{eqnarray}
The integration measure for the metric (\ref{LTBmetric}) at the
$t=t_0$ hypersurface reads as
\begin{equation}\label{detgij}
\sqrt{{\rm det}[g_{ij}(t=t_0)]} d^3 x = r^2 \sin\theta dr d\theta d\varphi~.
\end{equation}

Now we have all the ingredients to apply the Buchert equations
(\ref{Buchert1}), (\ref{Buchert2}) and (\ref{Buchert3}) to the flat
LTB metric (\ref{LTBmetric}). For symmetry and simplicity, we only
consider averages over a spherical domain, denoted $\mathcal{R}$, of
radius $R$ centered at the origin. Hence, plugging Eqs.\
(\ref{shearnow}), (\ref{expansionnow}) and (\ref{detgij}) into the
expression of the backreaction (\ref{backreaction}) yields
\begin{eqnarray}\label{extrastep1}
\mathcal{Q}_{\mathcal{R}} (t_0) = \frac{2}{3} \bigg[ \frac{3}{R^3} \int_{0}^{R} ((3 H_0(r) + rH_0'(r))^2 r^2 dr+ \nonumber \\
- \frac{3}{R^3} \int_{0}^{R} (rH_0'(r))^{2}) r^2 dr - \left( \frac{3}{R^3} \int_{0}^{R} (3 H_0(r) + r H_0'(r) ) r^2 dr \right)^2 \bigg]~,
\end{eqnarray}
where we have also used the spatial average as defined in Eq.\
(\ref{average}). Rewriting the terms of the integrands in Eq.\
(\ref{extrastep1}) as total derivatives, gives
\begin{equation}\label{LTBbackreaction}
\mathcal{Q}_{\mathcal{R}} (t_0) = \frac{2}{3} \left[ \underbrace{ \frac{3}{R^3} \int_{0}^{R} \frac{d}{dr} ( 3 r^3 H_0^2(r) ) dr}_{ = 9 H_0^2(R)}
- \underbrace{\left( \frac{3}{R^3} \int_{0}^{R} \frac{d}{dr} (r^3 H_0(r)) dr \right)^2}_{ = 9 H_0^2(R)} \right]~.
\end{equation}

Since the reference time $t=t_0$ is completely general, we can
conclude on the grounds of Eq.\ (\ref{LTBbackreaction}), that
\begin{equation}\label{backreactionvanishes}
\mathcal{Q}_{\mathcal{R}} (t) = 0
\end{equation}
for any time coordinate $t$ and averaging radius $R$. Altogether, we
have verified that the backreaction $\mathcal{Q}_{\mathcal{R}}$
vanishes identically for the flat matter dominated LTB model, as was
already shown in \cite{Paranjape:2006cd}. As an aside, it does not
vanish for an integration domain of arbitrary shape; we have checked
numerically that for various profiles of $H_0(r)$, the backreaction
is nonzero e.g.\ over a cubic region.

Moreover, transforming the coordinates of the $t=t_0$ hypersurface
shows that the spatial metric
\begin{equation}\label{dsigma2}
d\sigma^2= \left(\frac{ \partial A_0}{\partial r}\right)^2 dr^2 + A_0^2(r) (d\theta^2+\sin^2\theta
d\varphi^2)=dA_0^2+A_0^2(d\theta^2+\sin^2\theta d\varphi^2)~,
\end{equation}
reduces to the flat Euclidean form, for which the Ricci scalar
vanishes and hence we have $\langle{}^{(3)}R\rangle_\mathcal{D}=0$
in Eq.\ (\ref{Buchert2}).

Since $\mathcal{Q}_{\mathcal{R}}=0$ and $\langle {}^{(3)}R
\rangle_{\mathcal{R}}=0$, the Buchert equations (\ref{Buchert1}),
(\ref{Buchert2}) and (\ref{Buchert3}) reduce to the corresponding
FRW equations for the averaged scale factor $a_\mathcal{R}(t)$ and
the averaged matter density $\langle\rho\rangle_\mathcal{R}$:
\begin{eqnarray}
\frac{\ddot{a}_\mathcal{R}}{a_\mathcal{R}} &=& - \frac{4 \pi G}{3} \langle\rho\rangle_\mathcal{R}~, \label{Buchert1LTB} \\
\left(\frac{\dot{a}_\mathcal{R}}{a_\mathcal{R}}\right)^2 &=& \frac{8 \pi
G}{3} \langle\rho\rangle_\mathcal{R}~, \label{Buchert2LTB} \\
\frac{\partial}{\partial t} \langle\rho\rangle_\mathcal{R}&=&-3\frac{\dot{a}_\mathcal{R}}{a_\mathcal{R}}\langle\rho\rangle_\mathcal{R}~.
\label{Buchert3LTB}
\end{eqnarray}
The only difference in the Eqs.\ (\ref{Buchert1LTB}),
(\ref{Buchert2LTB}) and (\ref{Buchert3LTB}) compared to the FRW
equations is the dependence on the averaging scale. Therefore, they
also have the Friedmann solution, $a_{\mathcal{R}}(t)=
(t/t_0(R))^{2/3}$ with $t_0(R)$ as the scale dependent age of the
universe, and the template metric (\ref{averagedmetric}) reduces to:
\begin{equation}\label{averagedmetricFRW}
ds^2 = -dt^2 + \left( t/t_0(R) \right)^{4/3} \left[ dr^2 + r^2 (d\theta^2 + \sin^2\theta d\varphi^2) \right]~.
\end{equation}

One can now apply the metric (\ref{averagedmetricFRW}) and the
averaged equations (\ref{Buchert1LTB}), (\ref{Buchert2LTB}),
(\ref{Buchert3LTB}) to calculate the distance-redshift relations
(\ref{lumdist}) and (\ref{angdist}). By evaluating the expectation
value of the expansion scalar,
\begin{equation}\label{thetascalefactor}
\langle\theta\rangle_\mathcal{D}=\frac{\int_{\mathcal{D}}
\partial_t \sqrt{{\rm det} [g_{ij}]}d^3x }{\int_{\mathcal{D}}
\sqrt{{\rm det}
[g_{ij}]}d^3x}=3\frac{\dot{a}_\mathcal{D}}{a_\mathcal{D}} ~,
\end{equation}
we obtain a simple relation between the averaged Hubble constant $
\mathcal{H}_\mathcal{R} (t_0) \equiv
\dot{a}_\mathcal{R}(t_0)/a_\mathcal{R}(t_0)$ and the LTB Hubble
function (\ref{Timedependenthubble}):
\begin{equation}\label{thetahubble}
\mathcal{H}_\mathcal{R} (t_0) =
\frac{1}{3}\langle\theta_0\rangle_\mathcal{R} =\langle H_0(r)
+\frac{1}{3} rH_0'(r)\rangle_{\mathcal{R}} = \frac{1}{R^3}
\int_{0}^{R} \frac{d}{dr} (r^3 H_0(r)) dr = H_0(R) ~,
\end{equation}
where we have also used Eq.\ (\ref{expansionnow}). Using
$A(r,t)=a_\mathcal{R}(t) r$ with Eqs.\ (\ref{dtdz}), (\ref{drdz}),
(\ref{Buchert2LTB}) and (\ref{Buchert3LTB}), we obtain the averaged
distance-redshift relation
\begin{equation}\label{buchertrz}
\bar{r}(z)= 2 H_0^{-1}(R) \bigg( 1-\frac{1}{\sqrt{1+z}} \bigg) ~,
\end{equation}
where the averaged Hubble constant $\mathcal{H}_\mathcal{R} (t_0)$
has been eliminated with the help of Eq.\ (\ref{thetahubble}).
Utilizing Eqs.\ (\ref{lumdist}), (\ref{angdist}) and
(\ref{buchertrz}), the observables can be written in the final form:
\begin{equation}\label{BuchertdL}
\bar{d}_L(z)=(1+z)^2 a_\mathcal{R} (\bar{t}(z)) \bar{r}(z)= 2 H_0^{-1}(R(z)) (1+z) \bigg( 1 - \frac{1}{\sqrt{1+z}} \bigg) ~,
\end{equation}
\begin{equation}\label{BuchertdA}
\bar{d}_A(z)= a_\mathcal{R} (\bar{t}(z)) \bar{r}(z)=\frac{2 H_0^{-1}(R(z))}{(1+z)} \bigg( 1 - \frac{1}{\sqrt{1+z}} \bigg) ~,
\end{equation}
where instead of a single scale $R$, we have allowed for a running
averaging scale $R(z)$. Although then the averaged equations (\ref{Buchert1LTB}),
(\ref{Buchert2LTB}) and (\ref{Buchert3LTB}) are not anymore satisfied identically,
the equations hold true for each scale $R(z)$ separately. In practice, this means that we
are considering a different FRW model for each redshift.

The consequences of the redshift dependent
averaging scale in the observable relations (\ref{BuchertdL}) and
(\ref{BuchertdA}) constitute the subject of the following section.

\section{Observables in scale dependent Buchert averaging}\label{scale}

The conventional way to apply the Buchert equations is to choose a
single averaging domain $\mathcal{D}$, appropriate for the physical
system in question. This approximation has been justified
qualitatively on the basis of the observed statistical homogeneity and
isotropy \cite{Rasanen:2006kp}: For example, when taking the
averages over an observer-centered ball of radius $R$, its size
would have to be large enough compared to the small scale lumpiness.
As long as this condition is fulfilled $-$ the argumentation goes
$-$ the actual value of $R$ would be irrelevant.

However, when applied to large scale inhomogeneities, the validity
to use only a single averaging domain has not been examined
carefully before, though there have been some hints towards the idea
of multiple averaging scales \cite{Buchert:2007ik,Paranjape:2006ww}.
In the following, we consider this
quantitatively with the explicit toy model of Sect.\
\ref{buchertLTB}, for which it was shown that after averaging, the
scale dependence remains the only difference from the homogeneous
and flat matter dominated FRW case.

To make the treatment physically reasonable, we calculate
observables, such as the luminosity-distance $\bar{d}_L(z)$ and the
angular-diameter distance $\bar{d}_A(z)$, given by the appropriate
averaged expressions (\ref{BuchertdL}) and (\ref{BuchertdA}). The
deviation from their exact counterparts (\ref{lumdist}) and
(\ref{angdist}) is used as a measure for the goodness of the
approximation. We proceed by simply stating the results in Sects.\
\ref{analytic}, \ref{bubble}, \ref{structure} and leave the
discussion of the consequences to Sect.\ \ref{discussion}.

\subsection{Four levels of coarse graining}\label{threecases}

In this section, we introduce various ways to employ coarse
graining, later used for each model in Sects.\ \ref{analytic},
\ref{bubble} and \ref{structure}. In all of the cases, the angular
diameter distance is related to the luminosity distance by the
trivial factor, $d_L(z) = (1+z)^2 d_A(z)$, so we only give $d_L(z)$
for the different cases. We list the cases here with their later
referred names in bold:

\begin{itemize}

\item No coarse graining $-$ \textbf{Exact LTB}

Here the observables are determined by the exact Eqs.\
(\ref{lumdist}) and (\ref{angdist}), which we use as a baseline for
the other cases. Conceptually, this model serves as a source of the
actual observations a creature living in the exact LTB universe
would make.

\item A single averaging scale $R$ $-$ \textbf{Single scale}

In this case, the observables are determined by the Eqs.\
(\ref{BuchertdL}) and (\ref{BuchertdA}) with $R(z) = {\rm{constant}}
\equiv R$:
\begin{equation}\label{dLsinglescale}
\bar{d}_L(z) = 2 H_0^{-1}\left(R\right) (1+z) \bigg( 1 - \frac{1}{\sqrt{1+z}} \bigg) ~.
\end{equation}
When considering a model that fits the supernova observations in
Sect.\ \ref{bubble}, we take the averaging scale $R$ as the
present-day physical distance to the object with the highest
redshift in the supernova sample,
$R=r_{{\rm{LTB}}}(z_{{\rm{max}}})$, numerically computed from Eqs.\
(\ref{dtdz}) and (\ref{drdz}). Other choices would at most
correspond to different values of the Hubble constant, as is evident
in Eq.\ (\ref{dLsinglescale}). For periodic inhomogeneities in
Sect.\ \ref{structure}, we choose $R=2 \pi r_0$, where $r_0$ is the
wavelength of the inhomogeneities.

\item A different averaging scale $R(z)$ for each redshift $z$ $-$ Running scale case.

We further divide this into two distinct subcases, according to the
explicit form of the function $R(z)$:

\begin{enumerate}

\item $R(z)=\bar{r}(z)$ $-$ \textbf{Running scale with averaged geodesics}

In this case, we take $R(z)$ as the present-day physical distance to
each redshift, determined by the averaged geodesics
(\ref{buchertrz}), in which we take
$R=r_{{\rm{LTB}}}(z_{{\rm{max}}})$ or $R=2 \pi r_0$. Choosing again
$R=\bar{r}(z)$ would lead to the iterative use of Eq.\
(\ref{buchertrz}), perhaps ultimately converging to $r_{{\rm{LTB}}}(z)$ and making it
no different from our next case. Although not used here, this could
be a practical way of computing the distance in more realistic
models where the exact result is unattainable. The observables are
determined by
\begin{equation}\label{dLrunningscale}
\bar{d}_L(z) = 2 H_0^{-1}\left(2 H_0^{-1}(R) \bigg( 1-\frac{1}{\sqrt{1+z}} \bigg)\right) (1+z) \bigg( 1 - \frac{1}{\sqrt{1+z}} \bigg) ~.
\end{equation}

\item $R(z)=r_{{\rm{LTB}}}(z)$ $-$ \textbf{Running scale with exact geodesics}

Here we take the present-day physical
distance to each object as the running averaging scale, determined by the geodesics of the exact
metric, Eqs.\ (\ref{dtdz}) and (\ref{drdz}), yielding for the observables:
\begin{equation}\label{dLrunningscaleLTB}
\bar{d}_L(z) = 2 H_0^{-1}\left(r_{{\rm{LTB}}}(z)\right) (1+z) \bigg( 1 - \frac{1}{\sqrt{1+z}} \bigg) ~.
\end{equation}
\end{enumerate}
\end{itemize}

\subsection{Analytic considerations $-$ small $z$ behavior}\label{analytic}

There are two complications in comparing different realizations of
the scale dependent averaging without specifying the boundary
condition function $H_0(r)$. Firstly, there does not exist
expressions for the exact observables (\ref{lumdist}) and
(\ref{angdist}) in terms of elementary functions. Secondly, already
the relative comparison of the expressions for the averaged
observables (\ref{dLsinglescale}), (\ref{dLrunningscale}) and
(\ref{dLrunningscaleLTB}) is unfeasible without knowing something
about the function $H_0(r)$.

A natural solution for the two problems is to calculate Taylor
expansions for both the coarse grained observables of Eqs.\
(\ref{dLsinglescale}), (\ref{dLrunningscale}),
(\ref{dLrunningscaleLTB}), and the exact expressions
(\ref{lumdist}) and (\ref{angdist}). It is straightforward to take
the expansions to any desired order, but as the expressions become
more complicated and unillustrative for higher orders, we give them
to third order in redshift $z$. The price to pay is that the
analytic comparison is valid only at small redshifts; we have
numerically tested that the expansions are usable up to redshifts $z
\sim 0.2$.

We list the expansions of the luminosity distance here for the
different cases following the entitling of Sect.\ \ref{threecases}:
\begin{itemize}
\item \textbf{Exact LTB}
\begin{eqnarray}\label{dLexact}
d_L(z) = H_0^{-1}(0) \bigg[ z + \left( \frac{1}{4} - \frac{H_0'(0)}{H_0^2(0)} \right) z^2 + \nonumber \\
\bigg( -\frac{1}{8} + \frac{1}{3} \frac{H_0'(0)}{H_0^2(0)} + 2 \frac{H_0'^2(0)}{H_0^4(0)} - \frac{1}{2} \frac{H_0''(0)}{H_0^3(0)} \bigg) z^3 + \mathcal{O} (z^4) \bigg]
\end{eqnarray}
\item $R(z)=R$ $-$ \textbf{Single scale}
\begin{equation}\label{dLsinglescaleEXP}
\bar{d}_L(z) = H_0^{-1}(R) \left[z + \frac{1}{4} z^2 - \frac{1}{8} z^3 + \mathcal{O} (z^4) \right]
\end{equation}
\item $R(z)=\bar{r}(z)$ $-$ \textbf{Running scale with averaged geodesics}
\begin{eqnarray}\label{dLrunningscaleEXP}
\bar{d}_L(z) = H_0^{-1}(0) \bigg[ z + \left( \frac{1}{4} - \frac{H_0'(0)}{H_0(0)H_0(R)} \right) z^2 + \nonumber \\
 \bigg( -\frac{1}{8} + \frac{1}{2} \frac{H_0'(0)}{H_0(0) H_0(R)} + \frac{H_0'^2(0)}{H_0^2(0) H_0^2(R)} - \frac{1}{2} \frac{H_0''(0)}{H_0(0)H_0^2(R)} \bigg) z^3 + \mathcal{O} (z^4) \bigg]
\end{eqnarray}
\item  $R(z)=r_{{\rm{LTB}}}(z)$ $-$ \textbf{Running scale with exact geodesics}
\begin{eqnarray}\label{dLrunningscaleLTBEXP}
\bar{d}_L(z) = H_0^{-1}(0) \bigg[ z + \left( \frac{1}{4} - \frac{H_0'(0)}{H_0^2(0)} \right) z^2 + \nonumber \\
\bigg( -\frac{1}{8} + \frac{1}{2} \frac{H_0'(0)}{H_0^2(0)} + 2 \frac{H_0'^2(0)}{H_0^4(0)} - \frac{1}{2} \frac{H_0''(0)}{H_0^3(0)} \bigg) z^3 + \mathcal{O} (z^4) \bigg]
\end{eqnarray}
\end{itemize}

\subsection{Acceleration without backreaction}\label{bubble}

The expansion of the dust dominated flat LTB universe can have
neither local nor averaged acceleration but nevertheless can, as
shown e.g.\ in Sect.\ 3.2 of Ref.\ \cite{Enqvist:2006cg}, fit the
supernova observations. Thus, from the observational point of view,
the model can have effective acceleration. Since the backreaction
vanishes in this model, the only possibility to account for the
effect within the Buchert averaging formalism seems to be the
running smoothing scale.

In this section, we use the inhomogeneity profile found in Sect.\
3.2 of Ref. \cite{Enqvist:2006cg}, that gives a good fit to the
Riess et.\ al.\ gold sample of 157 supernovae \cite{Riess:2004nr}.
The boundary condition function of this model is given by
\begin{equation}\label{bubblehubble}
H_0(r) = H + \Delta H e^{-r/r_0}~,
\end{equation}
where the parameters have the values $H + \Delta H = 65.5~{\rm{km/s/Mpc}}$,
$\Delta H = 16.8~{\rm{km/s/Mpc}}$ and $r_0=1400~{\rm{Mpc}}$.

\begin{figure}[!hbp]
\begin{center}
\includegraphics[width=15.0cm]{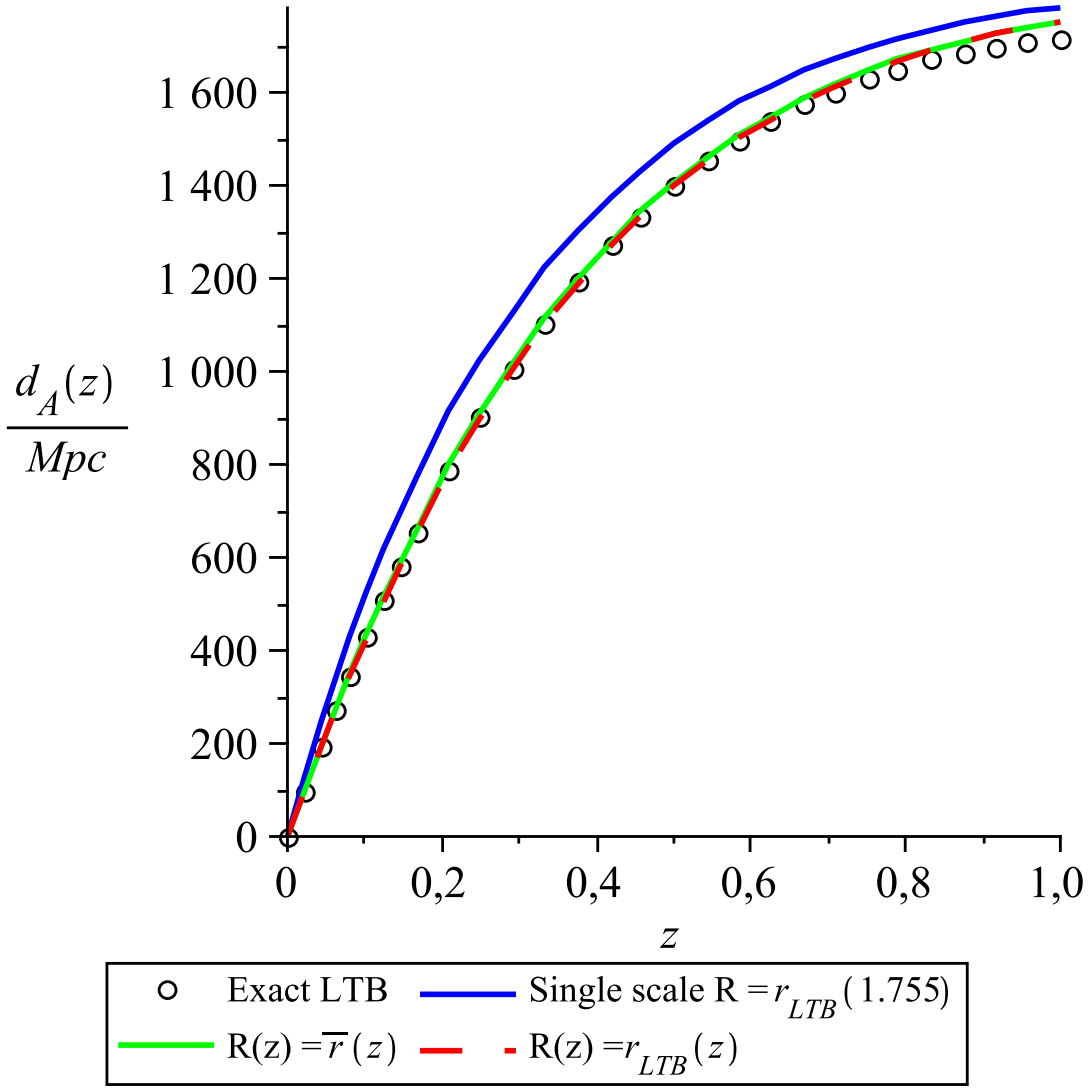}
\caption{The angular diameter distance $d_A(z)$ of the bubble model
with $H_0(r)=H+\Delta He^{-r/r_0}$, where $H+\Delta H=65.5 {\rm ~
km/s/Mpc}$, $\Delta H=16.8 {\rm ~ km/s/Mpc}$ and $r_0=1400 {\rm~
Mpc}$. The black circles represent the exact LTB solution, whereas
the blue, green and red curves correspond to the following coarse
grained cases: Blue $-$ a single averaging scale, chosen to be
$r_{{\rm LTB}}(1.755)=4944 {\rm ~ Mpc}$, where $z=1.755$ is the
redshift of the farthest supernova in the sample. Green $-$ the
averaged physical distance $\bar{r}(z)$ as the running smoothing
scale. Red $-$ the exact physical distance $r_{{\rm LTB}}(z)$ as the
running smoothing scale.} \label{bubbleabs}
\end{center}
\end{figure}

\begin{figure}[!hbp]
\begin{center}
\includegraphics[width=15.0cm]{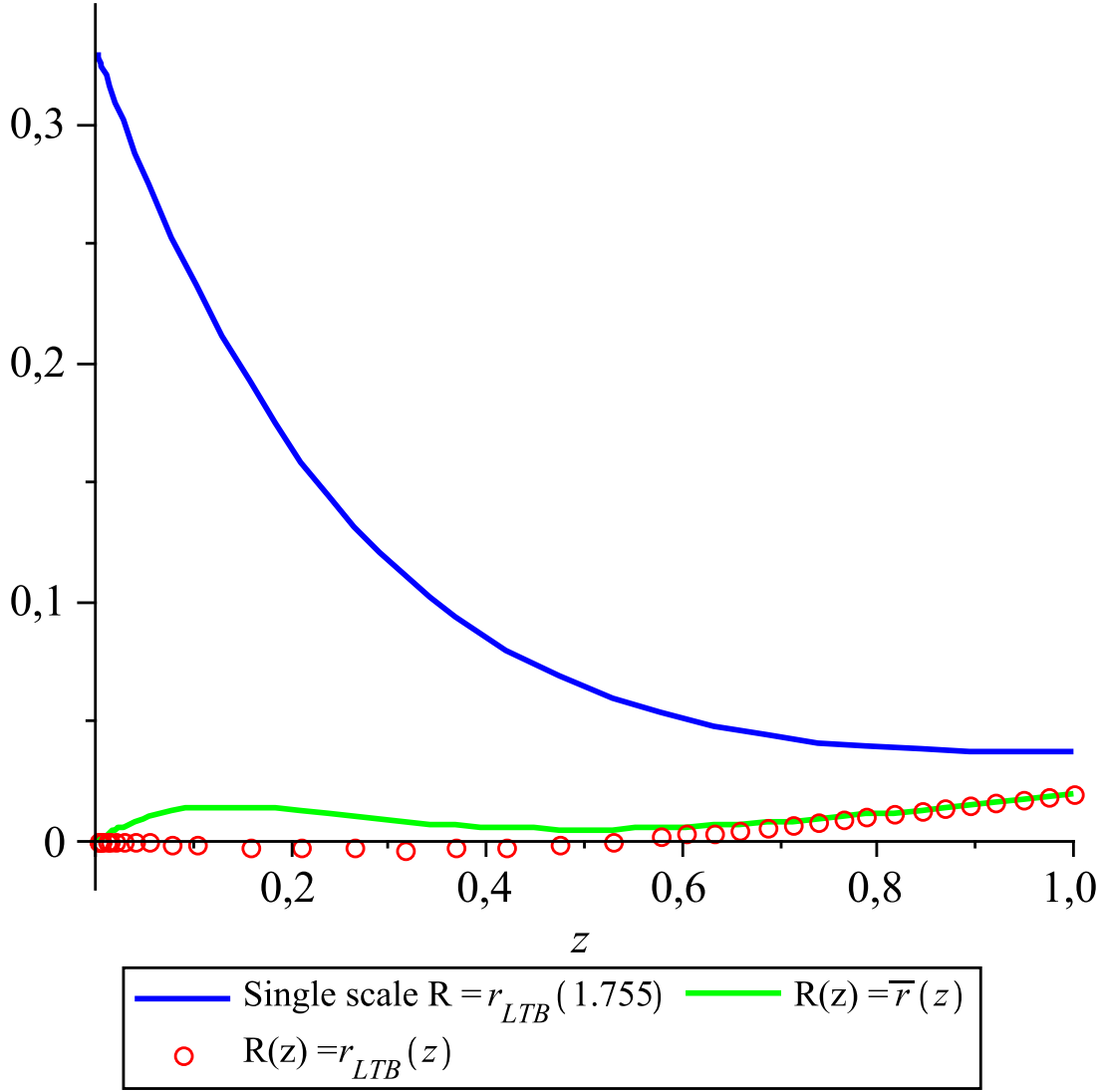}
\caption{The relative deviation $(\bar{d}_A(z)-d_A(z))/d_A(z) =
(\bar{d}_L(z)-d_L(z))/d_L(z)$ of the averaged angular diameter (or
luminosity) distance $\bar{d}_A(z)$ from the exact value $d_A(z)$
for the bubble model of Sect.\ 3.3 in the following cases: Blue $-$
a single averaging scale, chosen to be $r_{{\rm ~ LTB}}(1.755)=4944
{\rm ~ Mpc}$, where $z=1.755$ is the redshift of the farthest
supernova in the sample. Green $-$ the averaged physical distance
$\bar{r}(z)$ as the running smoothing scale. Red $-$ the exact
physical distance $r_{{\rm LTB}}(z)$ as the running smoothing
scale.}\label{bubblerel1}
\end{center}
\end{figure}

\begin{figure}[!hbp]
\begin{center}
\includegraphics[width=15.0cm]{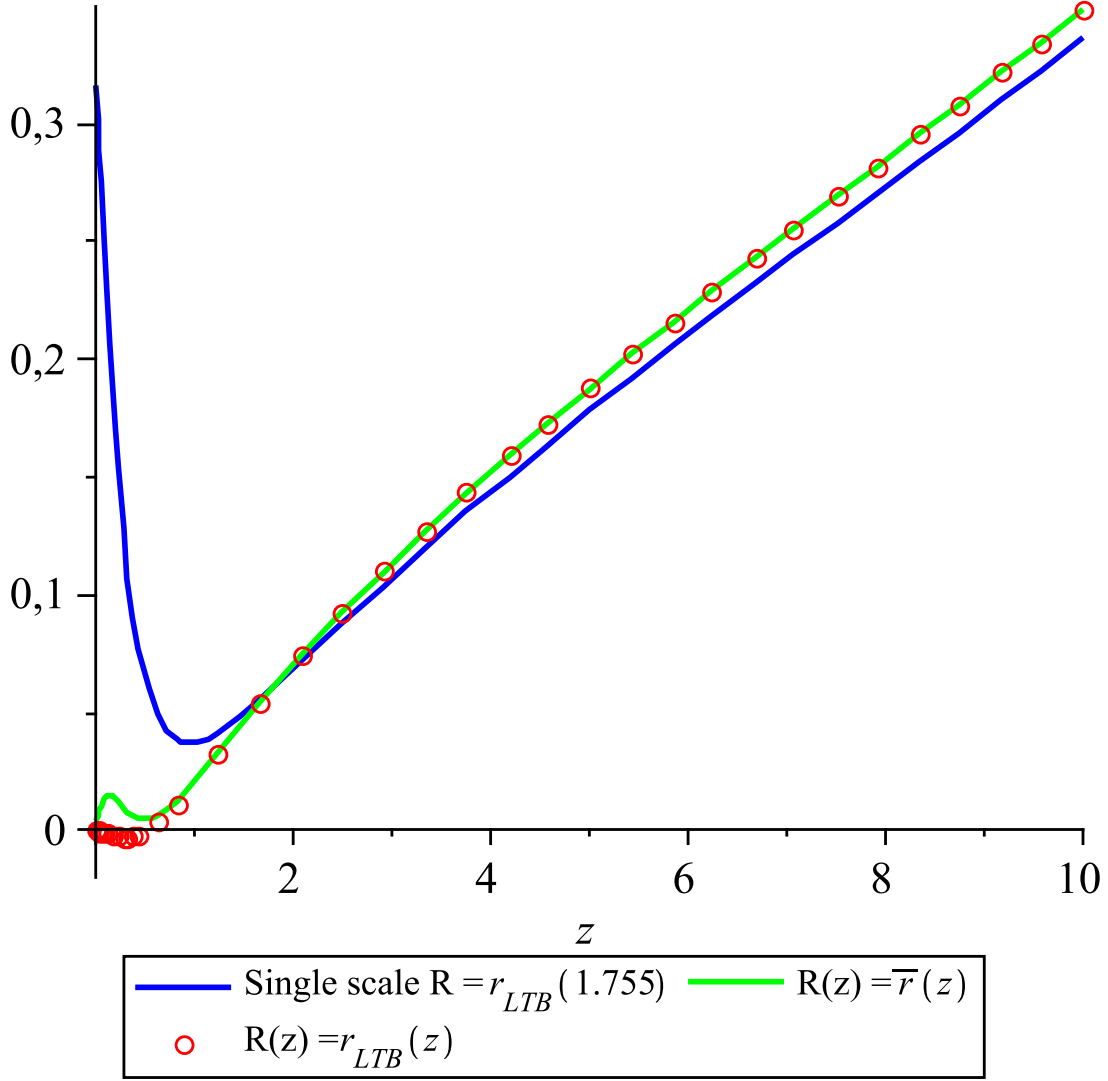}
\caption{Same as in Figure 2, but for larger redshift range,
$z=0...10$.}\label{bubblerel10}
\end{center}
\end{figure}

In Fig.\ \ref{bubbleabs}, we plot the exact angular diameter distance
of this model with the different averaged cases introduced in Sect.\
\ref{threecases}. Instead of the $d_L(z)$ observed from the
supernovae, we prefer to use $d_A(z)=d_L(z)/(1+z)^2$, as it is more
slowly increasing function of $z$, making the differences between
the various cases easier to detect. Moreover, the relative
deviations $(\bar{d}_A(z)-d_A(z))/d_A(z)$ are displayed in Figs.\ \ref{bubblerel1} and \ref{bubblerel10}, where $d_A(z)$ is the exact
result of Eq.\ (\ref{angdist}) and $\bar{d}_A(z)$ stands for the
averaged expressions (\ref{dLsinglescale}), (\ref{dLrunningscale}),
(\ref{dLrunningscaleLTB}). Due to the the general relation $d_L(z) =
(1+z)^2 d_A(z)$, the figures represent the relative deviations of
the luminosity distance as well.

Finally, we use the goodness of the fit to the Riess et.\ al.\
supernova data as an additional measure of the deviation, given by
\begin{equation}\label{chisquared}
\chi^{2} \equiv \frac{1}{157} \sum_{n=1}^{157} \left(
\frac{d_L^{{\rm{obs}}}(z_n)-d_L(z_n)}{\sigma_n} \right)^2~,
\end{equation}
where $\sigma_n$ is the estimated error of the measured luminosity
distance $d_L^{{\rm{obs}}}(z_n)$ to a source with redshift $z_n$. In
the different cases, $d_L(z)$ of Eq.\ (\ref{chisquared}) is
calculated from Eqs.\ (\ref{lumdist}), (\ref{dLsinglescale}),
(\ref{dLrunningscale}), (\ref{dLrunningscaleLTB}) and gives the
following values:
\begin{itemize}
\item \textbf{Exact LTB}: \begin{equation}\label{chiexact} \chi^2 = 1.12 \end{equation}
\item \textbf{Single scale} \begin{eqnarray}\label{chisingle} {\rm with}~R &=& 4944~ {\rm Mpc}:~\chi^2 = 4.35 \\
\label{chisingle2} {\rm with}~R &=& 905~ {\rm Mpc}:~\chi^2 = 2.06 \end{eqnarray}
\item \textbf{Running scale with averaged geodesics}: \begin{equation}\label{chirunning} \chi^2 = 1.11 \end{equation}
\item \textbf{Running scale with exact geodesics}: \begin{equation}\label{chirunningLTB} \chi^2 = 1.11 \end{equation}
\end{itemize}

\newpage

\newpage

\newpage

\vspace{-0.2cm}
\subsection{Periodic inhomogeneities as a toy model for structure}\label{structure}

Perhaps the closest representative of structure under the assumption
of spherical symmetry is achieved with a periodic boundary condition
function $H_0(r)$ \cite{Biswas:2006ub}. Hence, we take
\begin{equation}\label{structurehubble}
H_0(r) = H + \Delta H \sin{r/r_0}~,
\end{equation}
where the values $H = 65.5~{\rm{km/s/Mpc}}$, $\Delta H = 1.64 ~
{\rm{km/s/Mpc}}$ and $r_0=95 ~ {\rm{Mpc}}$ have been chosen to make
the plots as illustrative as possible. We do not consider more
intense inhomogeneities in order to keep the relation
$r_{{\rm{LTB}}}(z)$ single-valued to the redshift $z=1$.

\begin{figure}[!hbp]
\begin{center}
\includegraphics[width=15.0cm]{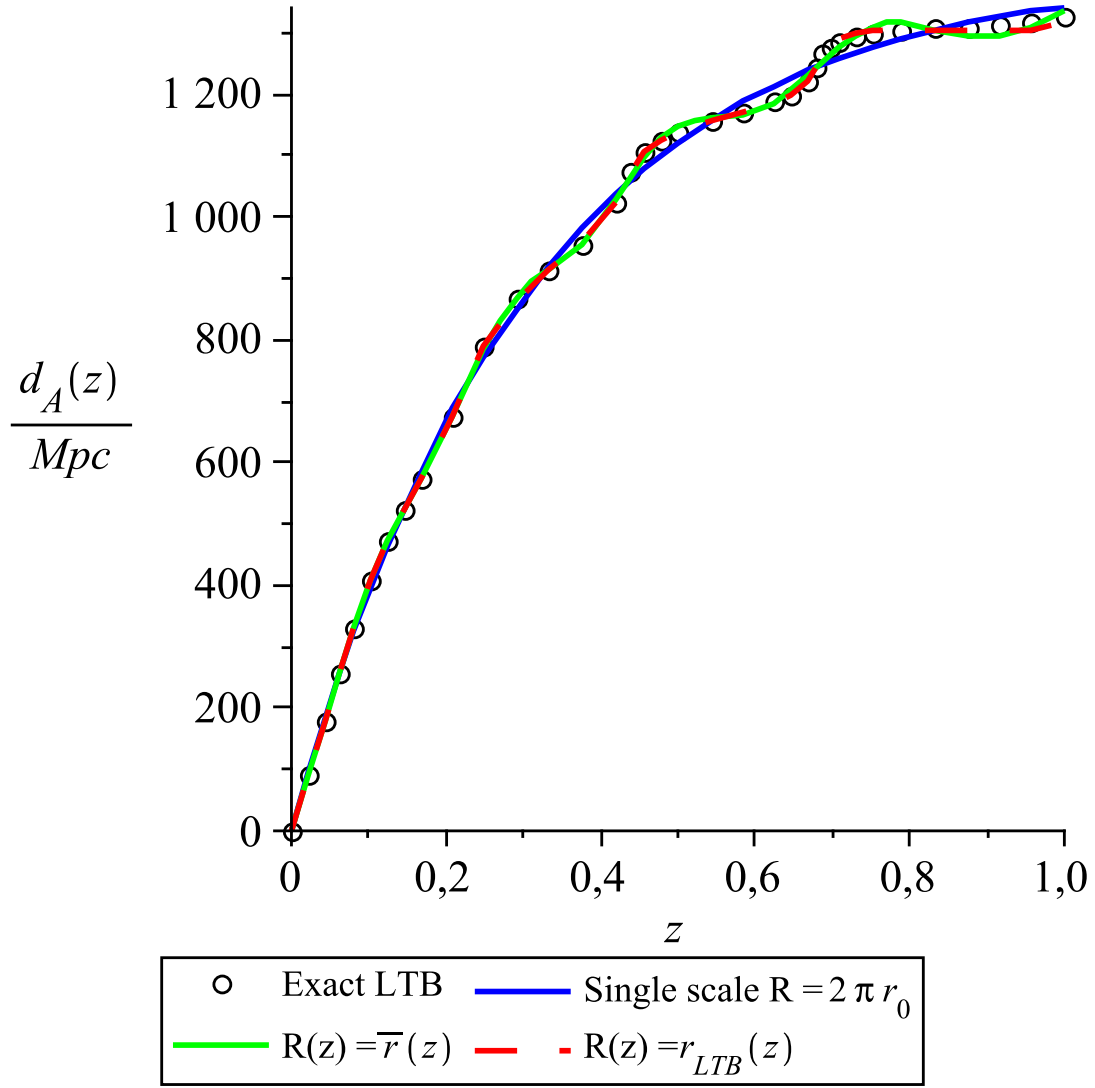}
\caption{The angular diameter distance $d_A(z)$ for the model of
Sect.\ 3.4 with periodic inhomogeneities, $H_0(r) = H + \Delta H
\sin(r/r_0)$, where $H=65.5 {\rm ~ km/s/Mpc}$, $\Delta H=1.64 {\rm ~
km/s/Mpc}$ and $r_0= 95 {\rm~ Mpc}$. The black circles represent the
exact LTB solution, whereas the blue, green and red curves
correspond to the following coarse grained cases: Blue $-$ a single
averaging scale, chosen to be one oscillation period $2 \pi r_0=598
{\rm ~ Mpc}$. Green $-$ the averaged physical distance $\bar{r}(z)$
as the running smoothing scale. Red $-$ the exact physical distance
$r_{{\rm LTB}}(z)$ as the running smoothing
scale.}\label{stuctureabs}
\end{center}
\end{figure}

\begin{figure}[!hbp]
\begin{center}
\includegraphics[width=15.0cm]{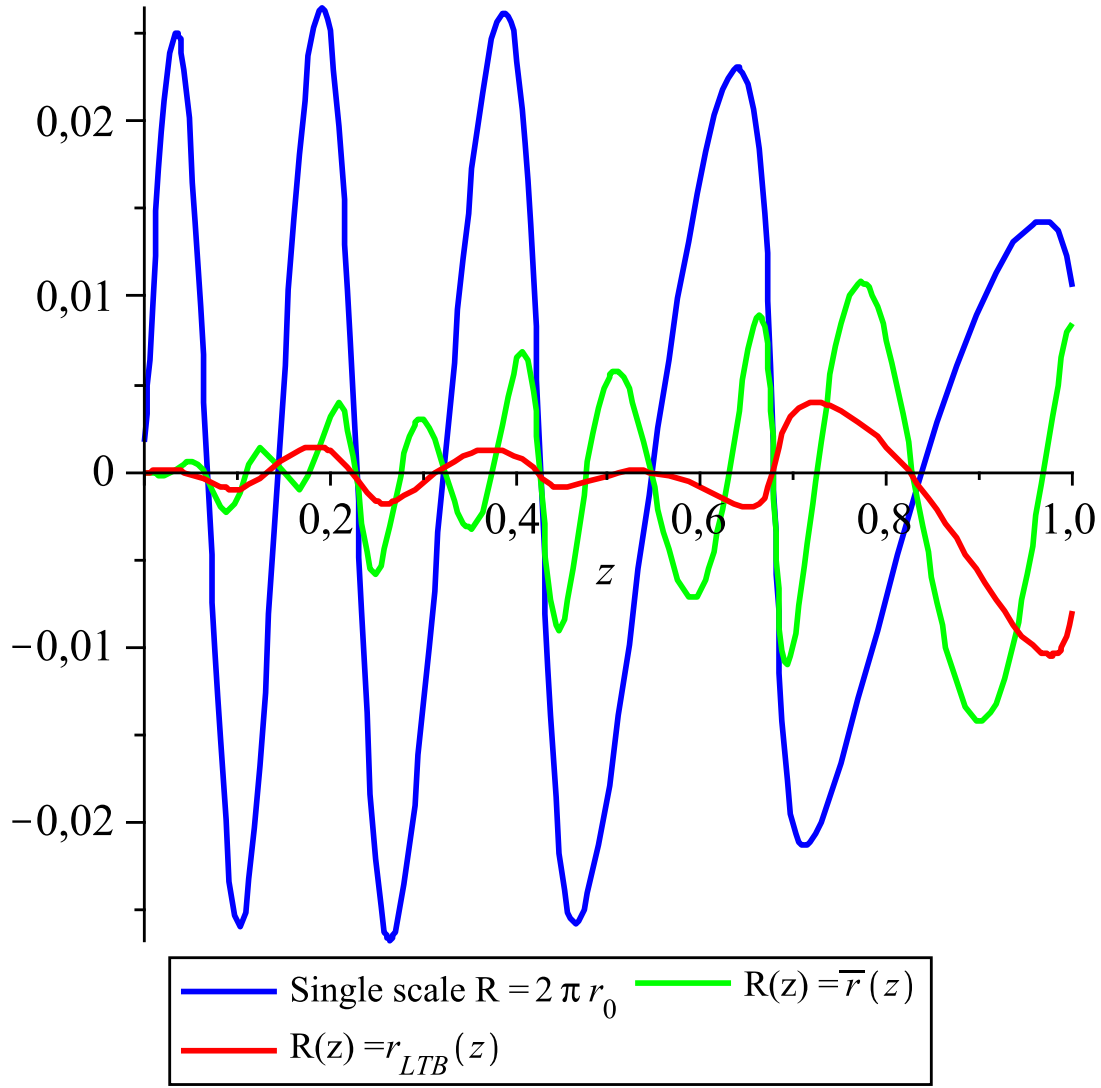}
\caption{The relative deviation $(\bar{d}_A(z)-d_A(z))/d_A(z) =
(\bar{d}_L(z)-d_L(z))/d_L(z)$ of the averaged angular diameter (or
luminosity) distance $\bar{d}_A(z)$ from the exact value $d_A(z)$
for the model of Sect.\ 3.4 in the following cases: Blue $-$ a
single averaging scale, chosen to be one oscillation period $2 \pi
r_0=598 {\rm ~ Mpc}$.  Green $-$ the averaged physical distance
$\bar{r}(z)$ as the running smoothing scale. Red $-$ the exact
physical distance $r_{{\rm LTB}}(z)$ as the running smoothing
scale.}\label{structurerel1}
\end{center}
\end{figure}

We plot the exact angular diameter distance of this model with the
averaged results in Fig.\ \ref{stuctureabs}, and the relative
deviation $(\bar{d}_A(z)-d_A(z))/d_A(z)$ in Fig.\
\ref{structurerel1}, for each of the different cases introduced in
Sect.\ \ref{threecases}.

\newpage

\newpage

\newpage

\subsection{Discussion of the results}\label{discussion}

Let us discuss here the results of Sects.\ \ref{analytic},
\ref{bubble} and \ref{structure} by taking each coarse graining
level of Sect.\ \ref{threecases} into separate consideration:

\subsubsection{$R(z)=R$ $-$ Single scale}

In this case, already the general form of the luminosity distance (\ref{dLsinglescale})
reveals the essential point: averaging
over a single scale is equivalent to using the perfectly homogeneous
and flat matter dominated FRW model. The freedom to choose the
averaging scale $R$ only corresponds to fixing the value of the
effective Hubble constant $H_0(R)$ in Eq.\ (\ref{dLsinglescale}).

The supernova data fit in the bubble model of Sect.\ \ref{bubble}
shows that when averaging over the domain that contains all the
supernovae in the sample ($R=4944~ {\rm Mpc}$), the resulting
$\chi^2 =  4.35$ represents a huge deviation from the exact result,
$\chi^2 = 1.12$. We get a better fit by choosing $R=905~{\rm{Mpc}}$,
because this gives the effective Hubble constant in Eq.\
(\ref{dLsinglescale}) its best fit value for the flat matter
dominated FRW case, $H_0(905~{\rm{Mpc}})=57.5~{\rm{km/s/Mpc}}$.
However, even this value gives $\chi^2 = 2.06$, which is still a way
too large deviation from the actual result. Besides, we have no a priori physical
justification to pick up the particular averaging scale
$R=905~{\rm{Mpc}}$. Overall, it is clear that
no single averaging scale can give an acceptable approximation for
the bubble model.

Perhaps the most interesting feature of the single scale case
becomes evident in Figs.\ \ref{stuctureabs} and \ref{structurerel1}:
even though the model of Sect.\ \ref{structure} with periodic
inhomogeneities is homogeneous on large scales, the single averaging
scale still leads to unwanted deviations. Whether it is an artifact
of the employed spherical symmetry, with light inevitably
propagating through all the layers of structure, or a more general
phenomenon, remains an open question.

Altogether, the expression of the averaged luminosity distance
(\ref{dLsinglescale}), along with its Taylor expansion
(\ref{dLsinglescaleEXP}), and the figures \ref{bubbleabs} -
\ref{structurerel1} all confirm the conclusion that averaging over a
single scale gives a too coarse-grained description at least for the
flat LTB universe. The inadequacy of the averaging procedure to
account for the observations in the LTB universe was already
suggested in Ref. \cite{Enqvist:2006cg}, but our results bring out
the essential point that the conclusion is valid only under the assumption of a
single smoothing scale, as we next discuss.

\subsubsection{$R(z)=\bar{r}(z)$ $-$ Running scale with averaged geodesics}

It has been speculated that averaging cosmological inhomogeneities
would be useful only when applied to systems with statistically
homogeneous distribution of small scale irregularities
\cite{Rasanen:2006kp}. Nevertheless, our results indicate that it is
possible to exploit averaging also for large scale inhomogeneities,
if the single smoothing scale $R$ is promoted to a redshift
dependent function $R(z)$.

On physical grounds, one could have expected some improvement in
accuracy of the observables when the single averaging scale $R$ is
replaced by the averaged present-day physical distance
(\ref{buchertrz}) to each object at redshift $z$. However, the
\emph{amount} of precision achieved with this generalization is both
surprising and a very welcome result. Indeed, the congruence between
this approximation and the exact results is evident in all of the
comparisons made in Sects.\ \ref{analytic}, \ref{bubble} and
\ref{structure} as we next specify in more detail.

When comparing the Taylor expanded luminosity distance
(\ref{dLrunningscaleEXP}) with the expansion of the exact LTB case
(\ref{dLexact}), one sees that to second order the results are almost
identical. They become exactly identical if we take $R=0$ as the
averaging scale for the geodesics, corresponding to the use of the
local Hubble parameter $H_0(0)$ in Eq.\ (\ref{buchertrz}), or if
$H_0(R)=H_0(0)$, as is the case for periodic inhomogeneities of
Sect.\ \ref{structure}. Moreover, even the third order terms carry
the same functional dependence on $H_0(r)$ as the exact case of Eq.\
(\ref{dLexact}), albeit with slightly different prefactors.

\vspace{-0.01cm}
The supernova data fit in the bubble model of Sect.\ \ref{bubble}
illustrates the power of the running scale approach in practical
applications. When employing the running scale, the goodness of the
fit changes from the puny $\chi^2=4.35$ of the single scale case to
the excellent fit $\chi^2=1.11$, which is within one percent of the
correct result $\chi^2=1.12$; the improvement is manifest in Figs.\ \ref{bubbleabs} and \ref{bubblerel1} as well. The result also
elucidates how the scale dependence can produce apparent
acceleration even in the absence of backreaction; for a thorough
discussion of the apparent acceleration, see page 9 of Ref.
\cite{Enqvist:2006cg}. Anyhow, in the real universe, one could
expect the effective acceleration to arise from the interplay
between scale dependence and backreaction, but the possibility
remains that either $-$ or neither for that matter $-$ of them plays
the dominant role.

\vspace{-0.01cm}
The toy model of structure in Sect.\ \ref{structure} also manifests
the advantage of the running smoothing scale: it is evident in
Figs.\ \ref{stuctureabs} and \ref{structurerel1} that the running scale
follows the oscillations of the exact observables whereas the single
scale case simply fails to.

\vspace{-0.01cm}
Problems arise, when going beyond the supernova fits to higher
redshifts. Indeed, Fig.\ \ref{bubblerel10} reveals that at $z \gtrsim 2$
the running scale falls short of the $\mathcal{O}(1\%)$
accuracy compared to the exact observables.
There is a plausible physical explanation for this: in the
employed LTB model, the growth of inhomogeneities \emph{backwards}
in time\footnote{See Eqs.\ (\ref{Timedependenthubble}),
(\ref{rhotime2}) and the paragraph thereafter.} makes them more
important at higher redshifts and cannot
be encapsulated in the present-day spatial averages. One can still
argue that in a more realistic model the problem would be
alleviated, since the inhomogeneities of the real universe are
expected to grow \emph{forwards} in time. Overall, we suppose the averaging with the running scale would be
conceivable at least up to $z \sim 2$; for higher
redshifts, one could then use coarser approximations, such as
the perturbed FRW models, presuming inhomogeneities really were of
less importance in the past.

\vspace{-0.01cm}
It would, nevertheless, be desirable to remedy the problems at higher redshifts.
For this, we have considered various forms for the running smoothing scale $R(z)$, beyond the
ones introduced in Sect.\ \ref{threecases}. The outcome is that there seems to be
no functions $R(z)$ that would both have a physical
basis, and improve the approximation over the whole range of
redshifts. In particular, it seems difficult to allow for the time evolution of the
universe within the running averaging scale approach.
Perhaps a better way to take into account the time evolution would be to average over
the past light cone. However, this would require a complete revision of the basic
formalism.

\vspace{-0.01cm}
Finally, a great virtue of using the geodesics of the averaged
metric is that one does not need to solve the exact geodesic
equations. Moreover, one can improve the accuracy of this method by
iterative use of the averaged distance-redshift relation
(\ref{buchertrz}), as argued in the paragraph before Eq.\
(\ref{dLrunningscale}).

\subsubsection{$R(z)=r_{{\rm{LTB}}}(z)$ $-$ Running scale with exact geodesics}

There are two conceptual steps in coarse graining needed to
calculate the observables \cite{Paranjape:2006ww}. Firstly, the step
from the exact Einstein equations to the Buchert equations, and
secondly, the step from the exact metric (\ref{LTBmetric}) to the
average metric (\ref{averagedmetricFRW}). To quantify the
approximation of the latter step, we studied a case where only
the field equations have been averaged, but instead of the average
metric, the exact metric determines the geodesics.

The outcome is that there are only minor deviations in the
observables between the use of the averaged and the exact geodesics.
This becomes apparent in the congruence between the Taylor
expansions of the luminosity distance (\ref{dLrunningscaleEXP}) and
(\ref{dLrunningscaleLTBEXP}), between the resulting $\chi^2$ for the
supernova data fit in Eqs.\ (\ref{chirunning}) and
(\ref{chirunningLTB}), and between the red and green curves in
Figs.\ \ref{bubbleabs} - \ref{structurerel1}. Although corresponding
only to a slight correction, it is still evident in all of these
comparisons that the exact physical distance gives the most accurate
approximation. Anyway, due to the good congruence of the results
between the averaged and the exact geodesics, the feasibility
becomes the deciding factor. Indeed, in more realistic models of the
universe, the exact geodesics are beyond computation. Overall,
perhaps the best solution in practice is to use the averaged
geodesics and, if needed, use Eq.\ (\ref{buchertrz}) iteratively as
explained in Sect.\ \ref{threecases}.

\section{Conclusions}\label{conclusions}

We have considered the role of scale dependence in the Buchert
averaging formalism, using the spherically symmetric and spatially
flat LTB dust universe as a testing ground. The vanishing of both
the backreaction and the spatial curvature scalar ${}^{(3)}R$ makes
this an ideal model to capture the effect of the averaging scale on
the observables, because then the only difference from the flat FRW
model is the explicit dependence on the smoothing scale. From the
resulting Buchert equations (\ref{Buchert1LTB}),
(\ref{Buchert2LTB}), (\ref{Buchert3LTB}) and the average metric
(\ref{averagedmetricFRW}), we have derived the luminosity and angular diameter distances
(\ref{BuchertdL}) and (\ref{BuchertdA}), carrying the scale dependence as well.
By employing a redshift dependent averaging scale $R(z)$, we have compared
these observables to the exact expressions (\ref{lumdist}) and
(\ref{angdist}). The physical reason to use a different averaging
scale for each object at redshift $z$ is clear: the distance the
observed light propagates depends on how far the object is.

Our principal result is that when the conventional single averaging
scale is replaced by the physical distance to each object at
redshift $z$, the relations (\ref{BuchertdL}) and (\ref{BuchertdA})
become significantly closer to the actual observables without
complicated computations. Although some improvement could be expected on
physical grounds, the ${\mathcal{O}}(1\%)$ precision at $z<2$ makes the
result a welcome surprise. Indeed, contrary to some previous
speculations \cite{Rasanen:2006kp,Enqvist:2006cg}, the result
suggests that the averaging procedure can be exploited for large
scale inhomogeneities, presuming the running smoothing scale is
employed. Although considered merely under the assumption of
spherical symmetry, we expect the running scale to show its full
advantage only when applied to more irregular large scale
inhomogeneities, such as the recently observed voids \cite{Rudnick:2007kw,Tikhonov:2007di}.
Naturally, the running scale can also be applied even if
a single scale would suffice, since then it reduces
to give the same predictions as the single scale approach.

By Taylor expanding the averaged observables and their exact
counterparts, we have demonstrated the increase in accuracy for generic
inhomogeneities up to redshifts $z \sim 0.2$. In addition, we have
numerically confirmed it for redshifts up to $z \sim 2$
using two explicit inhomogeneity profiles: a bubble inhomogeneity
that fits the supernova observations and periodic inhomogeneities as
a toy model for structure. Within the bubble model, we found that
the running smoothing scale can account for the apparent
acceleration even without backreaction. Consequently, it could be at
least as important as the backreaction in explaining dark energy as
an inhomogeneity induced illusion.

At redshifts $z\gtrsim 2$, the method breaks down. The plausible
physical explanation for this is that the present-day spatial
averages cannot capture the time evolution of the exact model.
Whether the problem can be solved within the already developed
formalism, or a generalized approach such as averaging over the
light cone would be needed, is something we hope to address in a
future work. In any case, the time evolution of the flat LTB model
is contrary to the observed structure formation, since the inhomogeneities
of the model grow towards large
redshifts. Hence, perhaps in the real universe one could apply the
running scale in averaging at low redshifts, and at higher
redshifts, resort to the conventional perturbed FRW models.

We have employed two ways to compute the present-day physical
distance $R(z)$ to objects with redshift $z$: firstly, using the
geodesics of the average metric and, for comparison, using the
geodesics of the exact metric. The outcome is that using the exact
geodesics gives only a marginal correction compared to the averaged
geodesics and, requiring the exact solution, would in general lead
to complicated computations. Overall, maybe the best method in practice
would be an iterative use of the averaged distance-redshift relation
(\ref{buchertrz}), as explained in Sect.\ \ref{threecases}.

Finally, there are problems in cosmological averaging we have not
addressed in this work (see
\cite{Rasanen:2006kp,Ellis:2005uz,Buchert:2007ik,Paranjape:2007wr,Celerier:2007jc,Coley:2007gw,Brouzakis:2006dj,Brouzakis:2007zi}).
Perhaps the most notable one within the Buchert formalism is that the averaged
equations (\ref{Buchert1}), (\ref{Buchert2}) and (\ref{Buchert3})
contain three equations for four unknowns. Therefore, more information is
in general needed to solve the equations. Whether this makes the
whole approach impractical in calculating observables outside the exact
solutions of the Einstein equations, is an open question.

\acknowledgments{We thank Tomi Koivisto, Kari Enqvist and Aseem Paranjape for helpful comments.
TM is supported by the Magnus Ehrnrooth Foundation. This work was also
supported by the European Union through the Marie Curie Research and
Training Network ``UniverseNet'' (MRTN-CT-2006-035863).}

\end{document}